\documentclass[preprint]{revtex4-1}

\usepackage{amsmath,amssymb}    
\usepackage{graphicx}   
\usepackage{verbatim}   
\usepackage{color}      
\usepackage{subfigure}  
\usepackage{hyperref}   
\hypersetup{colorlinks, linkcolor={red!50!black}, citecolor={blue!50!black}, urlcolor={blue!80!black}}
\usepackage{tikz}
\usetikzlibrary{decorations.pathmorphing}
\usepackage{lineno,hyperref}
\usepackage[utf8]{inputenc}
\usepackage{booktabs}
\usepackage{multirow}
\usepackage{siunitx}
\usepackage{calc}
\usepackage{color}
\usepackage{ulem}

\DeclareMathOperator{\arccot}{arccot}

\renewcommand{\Re}{\operatorname{Re}}
\renewcommand{\Im}{\operatorname{Im}}
\def\be{\begin{eqnarray}}
\def\ee{\end{eqnarray}}
\newcommand*{\red}{\textcolor{red}}

\begin{document}

\title{Three-dimensional black holes with quintessence}

\author{Jeferson de Oliveira}\email{jeferson@gravitacao.org}
\affiliation{Instituto de Física, Universidade Federal de Mato Grosso, CEP 78060-900, Cuiab\'a, Brazil}

\author{R. D. B. Fontana}\email{rodrigo.fontana@uffs.edu.br}
\affiliation{Universidade Federal da Fronteira Sul, Campus Chapec\'o, CEP 89802-112, SC, Brazil}

\date{}

\begin{abstract}

We study a quintessential black hole solution in three dimensions, with mass and quintessence charge.  By exploring the Carter-Penrose diagram, we show the presence of spacelike and lightlike singularities in the metric, given different values for the quintessence parameter, as well as an AdS-like spatial infinity and event horizon encapsulating the singularity. We also study the propagation of scalar and Dirac (Weyl) fields around the black hole solutions with different quintessence charges obtaining the quasinormal spectra for both fields using two different numerical methods with good agreement between the data. In both cases, the presence of quintessence increases the imaginary part of the quasinormal mode, since this is related to the event horizon of the solution, preserving the interpretation of this quantity as relaxation time in the corresponding CFT. We also investigate the behavior of high-temperature scalar field modes, demonstrating the presence of the so-called hydrodynamical limit, different from the BTZ black hole, for which no such modes exist.
\end{abstract} 

\maketitle

\section{Introduction}

The accelerated expansion of the Universe, discovered through type-Ia supernovae measurements \cite{Perlmutter:1998np,Riess:1998cb}, is nowadays a well-established fact . In the framework of general relativity, such an accelerated phase is expected to be driven by an exotic kind of matter with negative pressure called dark energy \cite{RevModPhys.75.559, Wang:2016lxa}. Besides the well-known models introducing this expansion by means of a cosmological constant term, some of the most promising candidates for dark energy models are those containing an extra scalar component, the quintessential field \cite{doi:10.1142/S021827180600942X}, introduced in the geometry via energy-momentum tensor. The effects of an accelerated expansion of the Universe can be probed by black hole physics, inasmuch as the dark energy field content will change the structure of spacetime. Since the 1990's a large number of works have studied black holes in models with a cosmological constant. In this work we deal with black holes solutions in an Universe with quintessence. 

In \cite{0264-9381-20-6-310}, Kiselev has found a family of static spherically symmetric black holes parameterized by the so-called quintessence charge - $w_q$ - and in \cite{Xu:2016jod} a generalization including rotation and charge was presented. In the context of AdS/CFT correspondence \cite{Maldacena:1997re,Witten:1998qj,1126-6708-2008-12-015}, Chen {\it{et al}}  \cite{Chen:2012mva} considered the effects of the quintessence field in the spacetime describing a $d$-dimensional planar AdS black hole. The authors found an exact solution with a planar topology depending on $w_q$, the cosmological constant $\Lambda$, and the black hole mass $M$. 

An important aspect that can be explored is the (in)stability of these black holes solutions against small perturbations to the neighborhood of event horizon via different probe fields. The solutions of the equations governing the evolution of the perturbations for "plane wave" boundary conditions are the so-called quasinormal modes. Such solutions have a characteristic spectrum of complex frequencies assumed the mentioned boundary conditions. For asymptotically flat black holes these are taken as purely ingoing waves at the horizon and purely outgoing modes at infinity, while for asymptotically AdS black holes, in general, the Dirichlet boundary condition is used \cite{Kokkotas:1999bd,Berti:2009kk,Konoplya:2011qq}.

Perturbations in black hole quintessential scenarios have been extensively studied in recent decades, considering the evolution of a variety of probe fields. The equations for Dirac, electromagnetic, and scalar fields with or without mass were analyzed in \cite{Chen:2005qh, Guo:2013mna, Ma:2006by, Zhang:2007nu,Zhang:2006hh,Varghese:2014xaa} for a quintessential Schwarzschild black hole, and their quasinormal modes were obtained. Additionally, in Reissner-Nordström quintessential scenarios, scalar, Dirac, and gravitational perturbations have been tested and the corresponding quasinormal spectra retrieved \cite{Varghese:2008ky, Wang:2009hr, Saleh:2009zz}. In general, the presence of a quintessential field in the metric yields oscillations with lower frequencies and damping factors. In the late-time behavior, the field is expected to decay exponentially, after the phase of quasinormal ringing \cite{Varghese:2014xaa}.


The evolution of a Gaussian wave package, representing a typical field perturbation in the neighborhood of a Schwarzschild black hole was studied by Vishveshwara \cite{Vishveshwara:1970zz} in the context of perturbation theory. This evolution comes in three different stages: the first characterized by a rapid initial pulse, given the initial burst of the perturbation; this is followed by a second stage of damped oscillations, called the quasinormal ringing phase; finally, the perturbations decay exponentially or as a power-law tail.

The question of stability of a given solution is answered by the analysis of the second and third phases of the field evolution. In the second phase, the sign of the imaginary part of the quasinormal frequencies settles whether the decay is stable. An initial perturbation should decay as a damped oscillation, for intermediate $t$ whenever $\Im(\omega)<0$; otherwise, the perturbation displays an unlimited growing mode and the system is unstable. In order to determine the field stability through the third phase of its evolution we must evolve the field for very long times and determine whether it displays a vanishing asymptotic behavior.

The study of AdS black holes in $d< 4$ spacetime dimensions is appealing in several ways \cite{PhysRevLett.69.1849} \cite{lemos2} \cite{Sa:1996ty}, as is the understanding of the dual field theory in the framework of AdS/CFT correspondence \cite{Son:2002sd} \cite{Abdalla:2011fd} and the study of the thermodynamical properties of such black holes, which are found to be similar to those of four-dimensional solutions, as shown for the BTZ black hole \cite{PhysRevD.47.3319} \cite{Brown:1994gs}.

In the context of AdS/CFT correspondence, the study of black hole quasinormal modes allows us to investigate specific aspects on the dual quantum field theories at finite temperature \cite{Maldacena:1997re} \cite{Witten:1998qj}. For instance, the inverse of the imaginary part of the fundamental quasinormal frequency is suggested to define a relaxation time scale for the dual thermal system to return to its equilibrium~\cite{Horowitz:1999jd}. In the case of quintessential AdS black holes, the computation of the quasinormal spectrum can bring some insight to a better understanding of the conjecture when the bulk has an accelerated expansion.

Our aim in this work is to present the causal structure of (2+1)-dimensional planar quintessential AdS black holes~\cite{Chen:2012mva} and also to study the evolution of classical fields in such geometry in order to obtain the quasinormal spectra, probing the stability of the metric to small field perturbations. The paper is organized as follows: In Sec.~\ref{solucao} we provide a brief review of the black hole solution considered in this paper. In Sec.~\ref{estruturacausal} we analyze its causal structure of and, in the following two sections, the propagation of scalar and Weyl fields and their quasinormal spectra. Section \ref{hidrodinamicos} brings the results on the calculation of quasinormal frequencies at high temperatures, after which, in Sec.~\ref{conclusoes}, we present the discussion of results and our conclusions.

\section{Three-dimensional Planar Black holes with quintessence}\label{solucao}


The line element ansatz for a $(2+1)$-dimensional planar black hole can be written as

\begin{equation}{\label{metrica1}}
ds^{2}=-A(r)dt^{2}+A(r)^{-1}dr^{2}+r^{2}dx^{2},
\end{equation}
where $r$ stands for the radial coordinate $0<r<\infty$ and $x$ is a planar coordinate $-\infty<x<\infty$. This is our starting point for the analysis of the geometrical properties of this geometry. 

The function  $A(r)$ is determined by Einstein's field equations,
\begin{equation}\label{einstein}
R_{ab}-\frac{1}{2}g_{ab}R-\frac{1}{L^{2}}g_{ab}=8\pi T_{ab},
\end{equation}
with $L$ standing for the AdS radius, related to the cosmological constant $\Lambda$ by $L^{2}=-1/\Lambda$, and $T_{\mu\nu }$ is the energy-momentum tensor for the quintessence. Following \cite{Chen:2012mva} \cite{0264-9381-20-6-310}, such a tensor can be cast in terms of the quintessence energy density $\rho_{q}$ and the state parameter of the quintessence, $w_q$, as
\begin{equation}\label{energymomentum}
T^{t}_{t}=T^{r}_{r}=-\rho_{q}, \hspace{0.3cm} T^{x}_{x}=(2w_q+1)\rho_{q}.
\end{equation}
Solving Einstein's field equations (\ref{einstein}) with the line element {\it{ansatz}} and the energy-momentum (\ref{energymomentum}), we have
\begin{equation}\label{solucaometrica}
ds^{2}=-\frac{r^{2}}{L^{2}}f(r)dt^{2}+\frac{L^{2}}{r^{2}}f(r)^{-1}dr^{2}+r^{2}dx^2,
\end{equation}
where 
\[
f(r)=1-\left(\frac{r_{+}}{r}\right)^{\sigma}, \hspace{0.3cm}  \sigma=2(1+w_{q}),
\]
and $r_{+}=\left(M L^2 \right)^{1/\sigma}$ specifies the event horizon, with $M$ being the black hole mass. As shown in Sec.~\ref{estruturacausal}, the solution (\ref{solucaometrica}) describes a class of black holes whose spatial infinity has the conformal structure of an AdS spacetime whatsoever the quintessence parameter $w_q$. In another way, the singularity formed at the point $r=0$ has a different character depending on the value of $w_q$, being the parameter crucial for the determination of the causal structure.

The line element in (\ref{solucaometrica}) can also be obtained, with the proper choice of charges, as a particular solution from a generic description of quintessential black holes by Kiselev \cite{0264-9381-20-6-310}. In particular, for $w_q=0$ we recover the metric of the well-known BTZ black hole \cite{PhysRevLett.69.1849}.

In the next section we clarify the coordinate singularity at $r\rightarrow r_+$ as a lightlike structure not singular in geometry, and establish the character of the singularity for different $w_q$.

\section{Causal Structure}\label{estruturacausal}
We now describe the causal structure for three different black hole solutions, corresponding to $w_q=-1/3$, $w_q=-2/3$ and $w_q=-1/2$. 

The spatial asymptotic form of the metric (\ref{solucaometrica}) does not depend on the quintessence parameter $w_{q}$. For all cases in the range $-1<w_{q}<0$, the spatial infinite $r\rightarrow \infty$ is AdS. The behavior of the  Kretschmann invariant, computed for the metric (\ref{solucaometrica}) shows
\begin{equation}\label{kretsch}
R_{abcd}R^{abcd}=\frac{12}{L^4}+\frac{1}{L^4}\left[p_1\left(\frac{r_{+}}{r}\right)^{2\sigma}-4p_2\left(\frac{r_{+}}{r}\right)^{\sigma}\right],
\end{equation}
 where $p_1=\sigma^{4}-6\sigma^3+15\sigma^2-20\sigma +12$ and $p_2=(\sigma-2)(\sigma-3)$. In the limit $r\rightarrow 0$,
 \[
 R_{abcd}R^{abcd}\rightarrow \infty,
 \]
 and for $r\rightarrow r_{+}$,
 \[
 R_{abcd}R^{abcd}\rightarrow \frac{1}{L^4}\left(\sigma^{4}-6\sigma^{3}+11\sigma^2\right).
 \]

Thus, the solution has a physical singularity at $r=0$ and the Kretschman invariant is well behaved at the location of the event horizon $r=r_{+}$ and approaches the AdS value $12/L^4$ at spatial infinity. In the following, we consider specific values of $w_q$ and go through the conformal diagrams, with the usual coordinate transformations.
 
\subsection{Black hole with $w_{q}=-1/3$} \label{A}

\begin{figure}[htp!]
\begin{center}
\includegraphics[scale=1.0]{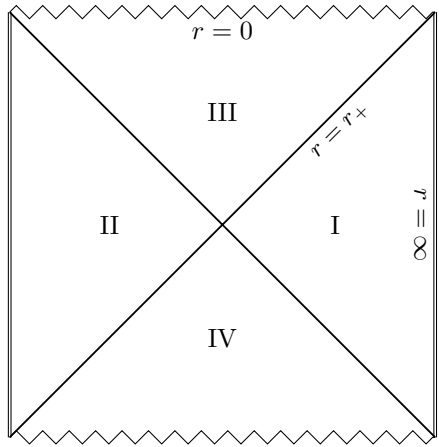}
\end{center}
\caption{Penrose-Carter diagram for the black hole with $w_{q}=-1/3$.} \label{penrose_1_3}
\end{figure}

In this case, the line element (\ref{solucaometrica}) takes the form
\[
 ds^{2}=-\frac{r^{2}}{L^{2}}\left[1-\left(\frac{r_{+}}{r}\right)^{4/3}\right]dt^{2}+\frac{L^2}{r^2\left[1-\left(\frac{r_{+}}{r}\right)^{4/3}\right]}dr^2+r^2dx^2.
 \]
 The first step in order to obtain the Penrose-Carter diagram of the black hole solution  is to define the advanced and retarded null coordinates $(u,v)$ as $u=t-r_{*}$ and $v=t+r_{*}$, respectively, where $r_{*}$ is the tortoise coordinate given by
\begin{equation}\label{tortoise_1}
r_{*}=\frac{3L^{2}}{2r_{+}}\left[\arccot\left(\frac{r}{r_{+}}\right)^{1/3}+\frac{1}{2}\log{\left(\frac{r^{1/3}-r_{+}^{1/3}}{r^{1/3}+r_{+}^{1/3}}\right)}\right],
\end{equation}
with $r_{*}\in ]-\infty,0]$. In this coordinate system, the metric is free of singularities at the event horizon. The maximal analytical extension of the metric is done by the Kruskal coordinates $U$ and $V$,
\begin{equation}\label{kruskal_1}
U=-e^{-\frac{2ur_{+}}{3L^2}-\pi/2}, \hspace{0.3cm} V=e^{\frac{2vr_{+}}{3L^2}-\pi/2},
\end{equation}
resulting in
\begin{equation}\label{uv_1}
-UV=e^{\left[2\arccot{\left(\frac{r}{r_{+}}\right)^{1/3}}-\pi\right]}\left(\frac{r^{1/3}-r_{+}^{1/3}}{r^{1/3}+r_{+}^{1/3}}\right).
\end{equation}
Defining the new coordinates $\tilde{U}=\arctan(U)$ and $\tilde{V}=\arctan(V),$ we have the Penrose-Carter diagram in Fig.\ref{penrose_1_3}. Notice that the spatial infinity is conformally AdS and the spacelike  singularity is located at $r=0$, being covered by the lightlike structure at $r=r_{+}$ (event horizon).

\subsection{Black hole with $w_{q}=-1/2$}

\begin{figure}[htp!]
\begin{center}
\includegraphics[scale=0.8]{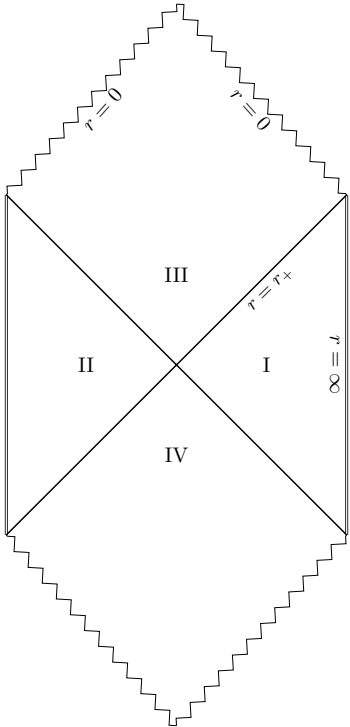}
\end{center}
\caption{Penrose-Carter diagram for the black hole with $w_{q}=-1/2$ and $w_{q}=-2/3$.} \label{penrose_light}
\end{figure}

Following the same steps as for the case $w_{q}=-1/3$, we find that the tortoise coordinate is given by
\begin{equation}\label{tortoise_2}
r_{*}=\frac{L^2}{r_{+}}\log{\left(1-\frac{r_+}{r}\right)}
\end{equation}
and the Kruskal coordinates are $U=-e^{-ur_{+}/2L^{2}}$ and $V=e^{vr_{+}/2L^{2}}$. Thus,
\[
-UV=\left(1-\frac{r_{+}}{r}\right).
\]
 Using the same Penrose coordinates $\tilde{U}$ and $\tilde{V}$ as in Sec.~\ref{A}, we can construct the Penrose-Carter diagram (Fig. \ref{penrose_light}) for this case. The surprising feature here is the change in the nature of the singularity: by decreasing $w_q$, we go from a spacelike singularity to a lightlike one.

The spacetime is still encapsulated by an event horizon at $r=r_+$, preventing the existence of a naked singularity.

\subsection{Black hole with $w_{q}=-2/3$}

Similarly to the previous case (Fig.~\ref{penrose_light}), the Penrose-Carter Diagram for the black hole with $w_{q}=-2/3$ also shows a lightlike spacetime singularity. For completeness, we present the tortoise coordinate and the Kruskal extension in this case as well:
\begin{equation}\label{tortoise_3}
r_{*}= \frac{3L^2}{r_{+}^{2/3}r^{1/3}}-\frac{3L^2}{2r_{+}}\log{\left[\frac{r^{1/3}+r_{+}^{1/3}}{r^{1/3}-r_{+}^{1/3}}\right]},
\end{equation}
\[
U=-e^{-ur_{+}/3L^{2}}, \hspace{0.3cm} V=e^{vr_{+}/3L^{2}}.
\]
Thus, 
\[
-UV=e^{2\frac{ r_{+}^{1/3}}{{r^{1/3}}}}\left[\frac{r^{1/3}-r_{+}^{1/3}}{r^{1/3}+r_{+}^{1/3}}\right].
\]
\\

\subsubsection{Singularity character: threshold for $w_{q}$}

The general tortoise coordinate can be written as a first-order hypergeometric function,
\be
\label{gentort}
r_*= L^2 \int \frac{1}{r^2 - r_+^\sigma r^{2-\sigma}}dr = \frac{ L^2 r^{1 + 2w_q} { _2}F_1 \left[ 1 , \frac{1+2w_q}{2+2w_q} , 1 + \frac{1+2w_q}{2+2w_q}, \frac{r^{2+2w_q}}{r_+^\sigma}\right]}{r_+^\sigma (-1-2w_q)}.
\ee
The limit $r \rightarrow 0$ lead to ${ _2}F_1 \rightarrow 1$, which allows us to investigate the threshold between the spacelike and the lightlike singularities. 
When $w_q > -1/2$, we have
\be
\label{gt2}
\lim_{r\rightarrow 0} r_* \rightarrow limited
\ee
which leads to
\be
\label{gt2}
\lim_{r\rightarrow 0} UV \rightarrow 1.
\ee
With the usual choice of $\tilde{U}$ and $\tilde{V}$ as in Sec.~A, we will always have a spacelike singularity. 

On the other hand, when $w_q \leq -1/2$, we will have
\be
\label{gt2}
\lim_{r\rightarrow 0} r_* \rightarrow \infty
\ee
and
\be
\label{gt2}
\lim_{r\rightarrow 0} UV \rightarrow \infty.
\ee 
what maps $\tilde{U}$ and $\tilde{V}$  into the points $\pm \pi /2$, generating a lightlike singularity, as in Secs. B and C. 

In spite of the topological difference between spacetimes for different $\sigma$, the singularities are always enclosed by an event horizon. This allows us to infer properties like the Hawking temperature in each solution and the usual thermodynamics coming from it, as well as to evolve dynamical fields, studying the quasinormal spectra of the solutions in the region outside the horizon. In order to obtain these spectra, we impose the usual boundary conditions to AdS-like spacetimes, consisting of plane waves entering the event horizon, and a vanishing field at spatial infinity. In the next sections we calculate the quasinormal modes for two different fields, analyzing the scaling of these modes with the black hole constants.  

\section{Klein-Gordon field perturbation}\label{kg}

The evolution of a massive probe scalar field $\Psi$ in the black hole geometry (\ref{solucaometrica}) is given by the well-known Klein-Gordon equation, which, after performing the separation of variables $\Psi(r,x,t)=\frac{Z(r)}{\sqrt{r}}e^{-i\omega t +i\kappa x}$, can be cast in the form

\begin{equation}\label{kg1}
\frac{d^{2}}{dr_{*}^{2}}Z(r)+\left[\omega^{2}-V(r)\right]Z(r)=0,
\end{equation}
where $V(r)$ is the effective potential. For the massless case $m^{2}=0$, $V(r)$ has the following expression

\begin{equation}\label{potencial_escalar}
V(r)=\frac{3r^{2}}{4L^{4}}-\frac{M^{2}}{4 r^{2(1+2w_{q})}}-\frac{M^{2}w_{q}}{r^{2(1+2w_{q})}}-\frac{M}{2L^{2}r^{2 w_{q}}}+\frac{\kappa^{2}}{L^{2}}+\frac{M w_{{q}}}{L^{2}r^{2 w_{{q}}}}-\frac{\kappa^{2}M}{r^{2(1+w_{q})}}.
\end{equation}

At the event horizon $r_{+}=(ML^2)^{\frac{1}{2(1+w_{q})}}$, the effective potential $V(r)$ goes to zero and is infinite at the conformal spatial infinity $r\rightarrow \infty$. In the limit of vanishing quintessence $w_{q}\rightarrow 0$, $V(r)$ is the effective potential of a massless scalar field evolving in a BTZ black hole~\cite{Cardoso:2001hn} as expected. In Fig.~(3) we show curves of $V(r)$ for different values of the quintessence parameter $w_{q}$. All curves display the same qualitative behavior, with the potential near the event horizon decreasing as the quintessence parameter approaches the pure AdS case $w_{q}=-1$. 

\begin{figure}[htp!]
\centering
\includegraphics[scale=1.3]{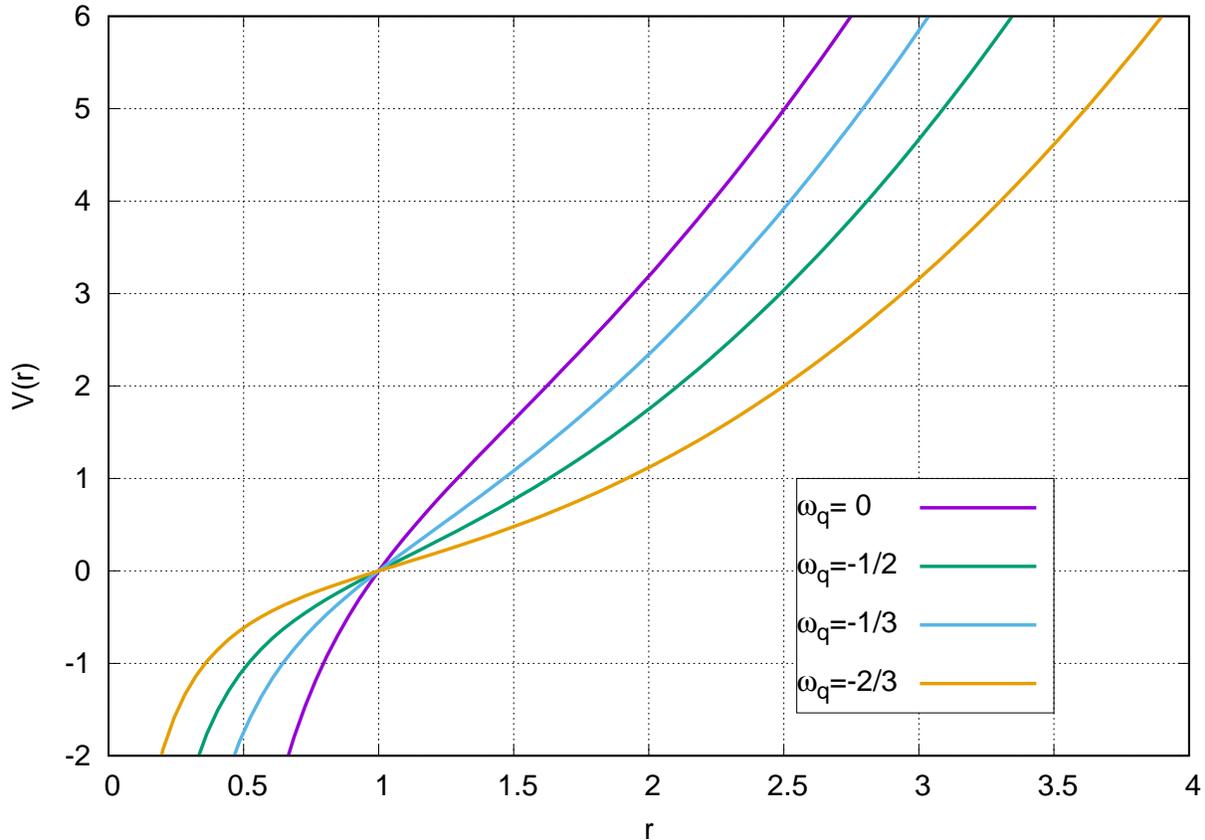}
\caption{Scalar field effective potential for several values of $w_{q}$ with $M=L=\kappa=1$.}
\label{pot_escalar}
\end{figure}

It is useful to compute the quasinormal spectrum due to the probe scalar field, since this allows us, e.g., to verify the black hole stability. Using e. g., the Horowitz-Hubeny technique \cite{Horowitz:1999jd} (in exact the same), for the effective potential (\ref{potencial_escalar}), we have obtained the frequencies. The results are listed in tables \ref{scalar_qnm__k_0_hh} and \ref{scalar_qnm__k_1_hh}. 
In the absence of quintessence, we reproduce the values for the BTZ black hole case \cite{Cardoso:2001hn}. Our data also reveal that, even when $w_q\neq 0$, the imaginary part of $\omega$ scales with the size of the black hole $r_{+}$, i.e., $-\Im(\omega)\simeq r_{+}$ remains valid while the scaling $\Re(\omega)\simeq \kappa$ is broken in the presence of the quintessence field.

The results of the tables \ref{scalar_qnm__k_0_hh} and \ref{scalar_qnm__k_1_hh} were also obtained with the characteristic integration in null coordinates  (along with the Prony method) as described in \cite{Konoplya:2011qq}, and we find both to be in very good agreement: the last digit for imaginary parts of $\omega$ in table \ref{scalar_qnm__k_0_hh} indicates the uncertainty inferred from the divergence between both methods (and for the cases $\Re(\omega) \neq 0$). The largest deviation we find is for the case $w_q=-\kappa /2=-0.5$ for which the numerical integration method yields $\omega = 0.2423-1.1517i$, a $0.4\%$ deviation from the value listed in the table.

\begin{table}[htbp]
\centering
  \begin{tabular}{lSSSSSSSS}
    \hline
    \multirow{3}{*}{$r_{+}$} &
      \multicolumn{2}{c}{$w_{q}=0$} &
      \multicolumn{2}{c}{$w_{q}=-0.8$} &
      \multicolumn{2}{c}{$w_{q}=-0.5$} &
      \multicolumn{2}{c}{$w_{q}=-0.1$}\\
      \midrule
      & {$\Re(\omega)$} & {$-\Im(\omega)$} & {$\Re(\omega)$} & {$-\Im(\omega)$} & {$\Re(\omega)$} & {$-\Im(\omega)$} & {$\Re(\omega)$} & {$-\Im(\omega)$} \\
      \hline
    1	& 0.0	& 2.00		& 0.0	& 0.249 	& 0.0	& 0.666	& 0.0	& 1.458	\\
    5	& 0.0	& 10.00		& 0.0	& 1.247		& 0.0	& 3.333 & 0.0	& 7.291		\\
    10	& 0.0	& 20.00		& 0.0	& 2.495		& 0.0	& 6.666 & 0.0	& 14.583		\\
    50	& 0.0	& 100.00	& 0.0	& 6.138		& 0.0	& 33.333& 0.0	& 72.916	\\
    100 & 0.0	& 200.00	& 0.0	& 12.477	& 0.0	& 66.667& 0.0	& 145.833	\\
    \hline
  \end{tabular}
   \caption{Scalar quasinormal frequencies for fixed $L=1$ and $\kappa=0$.}\label{scalar_qnm__k_0_hh}
\end{table}
\begin{table}[htbp]
\centering
  \begin{tabular}{lSSSSSSSS}
    \hline
    \multirow{3}{*}{$r_{+}$} &
      \multicolumn{2}{c}{$w_{q}=0$} &
      \multicolumn{2}{c}{$w_{q}=-0.8$} &
      \multicolumn{2}{c}{$w_{q}=-0.5$} &
      \multicolumn{2}{c}{$w_{q}=-0.1$}\\
      \midrule
      & {$\Re(\omega)$} & {$-\Im(\omega)$} & {$\Re(\omega)$} & {$-\Im(\omega)$} & {$\Re(\omega)$} & {$-\Im(\omega)$} & {$\Re(\omega)$} & {$-\Im(\omega)$} \\
      \hline
    1	& 1.00	& 2.00		& 0.1585	& 0.540 	& 0.241	& 1.153	& 0.794	& 1.896	\\
    5	& 1.00	& 10.00		& 0.0		                                                                                                                                                                                                                                                                                                                                                                                                                                                                                                                                                                                                                                                                                                                                                                                                                                                                                                                                                                                                                                                                                                                                                                                                                                                                                                                                                                                                                                                                                                                                                                                                                                                                                                                                                                                                                                                                                                                                                                                                                                                                                                                                                                                                                                                                                                                                                                                                                                                                                             & 1.301		& 0.0	& 3.409 & 0.0	& 7.509		\\
    10	& 1.00	& 20.00		& 0.0		& 2.521		& 0.0	& 6.704 & 0.0	& 14.688		\\
    50	& 1.00	& 100.00	& 0.0		& 12.483	& 0.0	& 33.341& 0.0	& 72.941	\\
    100 & 1.00	& 200.00	& 0.0		& 24.958	& 0.0	& 66.672& 0.0	& 145.853	\\
    \hline
  \end{tabular}
  \caption{Scalar quasinormal frequencies for fixed $L=1$ and $\kappa=1$.}\label{scalar_qnm__k_1_hh}
\end{table}

\begin{figure}
\label{figg1}
\resizebox{1.0 \linewidth}{!}{\includegraphics{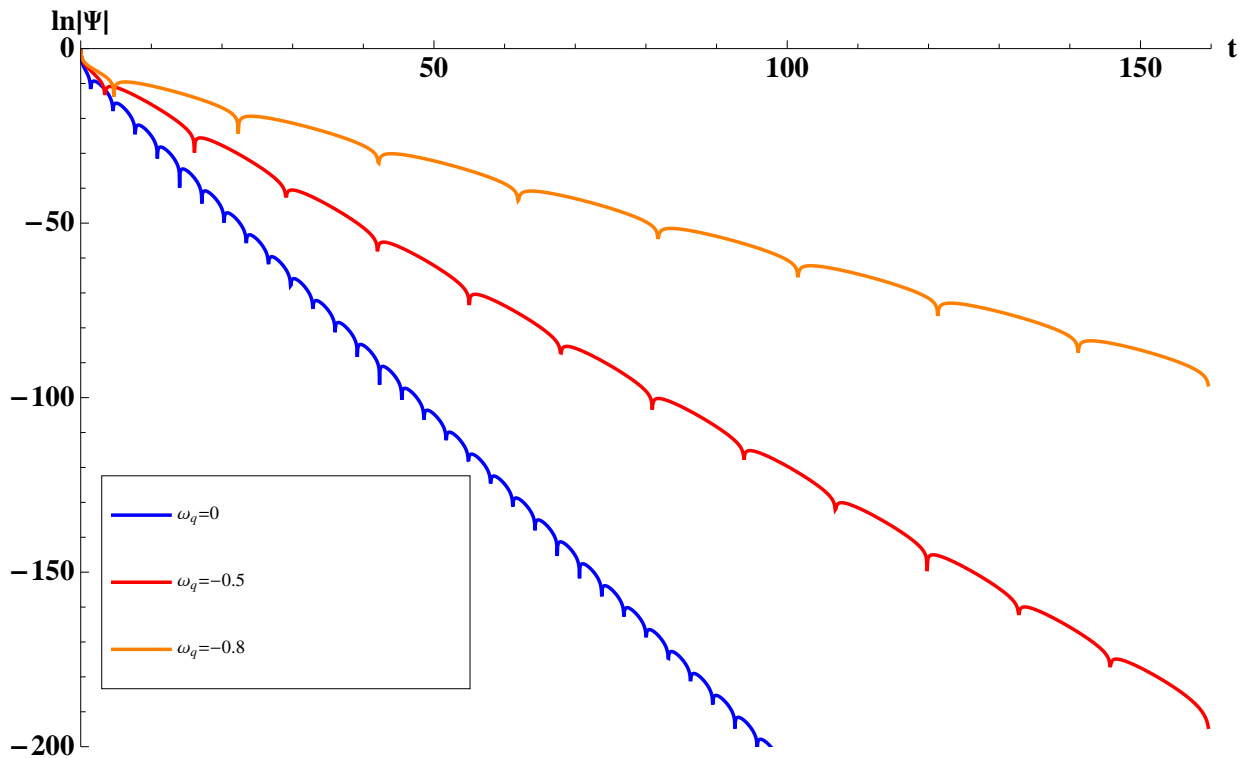}}
\caption{Scalar Field evolving in a BTZ-quintessential black hole. The geometry parameters read $r_+=L=\kappa=1$.}
\end{figure}

Examples of the time evolution of the field (obtained through the characteristic integration method), are shown in Fig. \ref{figg1}. We see that the imaginary and real parts of the fundamental mode decrease faster with increasing $|w_q|$, but the temporal evolution permanently remains that of a damped oscillator, not showing the transition to an exponential decay that others AdS-like spacetimes or different spin-fields can display. 

By the results showed in tables \ref{scalar_qnm__k_0_hh} and \ref{scalar_qnm__k_1_hh} it seems, a peculiar feature emerges for the quasinormal modes when $k>0$ (and $\sigma \neq 2$): for small $r_+$ the modes have $\Re(\omega)>0$, and for high $r_+$ they are purely imaginary. Given the possible qualitative change in $\omega$, we investigate this behavior for a specific case in the Appendix. We also address the question of a scale of $\omega$, $L$, $r_+$ and $k$, what is clear for $\sigma = 2$ ~\cite{Cardoso:2001hn}.

\section{Weyl field perturbation}\label{weyl}

We complete the study of field evolution by analyzing a massless fermion field dictated by the usual wave equation as found in \cite{Chandrasekhar:1985kt}. In order to study the dynamics of the Weyl field given by
\begin{equation}\label{dirac1}
i\gamma^{(a)}e_{(a)}^{\mu}\nabla_{\mu}\Psi-m\Psi=0,
\end{equation}
we consider the Dirac equation for a two-component massive spinor
\begin{equation}\label{two_spinor}
\Psi=\begin{bmatrix}
    \Psi_1(t,r,x) & \\
    \Psi_2(t,r,x) &
\end{bmatrix},
\end{equation}
where the indexes enclosed in parenthesis, $(a)$, refer to the flat coordinates in the tangent space, $\gamma^{(a)}$ are the usual gamma matrices, and $m$ denote the spinor mass. Following \cite{CuadrosMelgar:2011up} the spinor covariant derivative $\nabla_{\mu}$ can be written as

\begin{equation}\label{derivada_spinor}
\nabla_{\mu}=\partial_{\mu}+\frac{1}{8}\omega_{\mu}^{(a)(b)}\left[\gamma_{(a)},\gamma_{(b)}\right],
\end{equation}
where $\omega_{\mu}^{(a)(b)}$ denote  the the spin connection,
\begin{equation}\label{spin_connection}
\omega_{\mu}^{(a)(b)}=e_{\nu}^{(a)}\partial_{\mu}e^{(b)\nu}+e_{\nu}^{(a)}\Gamma^{\nu}_{\mu\alpha}e^{\alpha (b)}.
\end{equation}
The  triad $e_{(a)}^{\mu}$ and the metric connections $\Gamma^{\nu}_{\mu\sigma}$ for the line element (\ref{solucaometrica}) are given by the following expressions, where $g(r)=\frac{r^2}{L^2}f(r)$
\begin{equation}\label{triade}
e_{t}^{(a)}=\sqrt{g(r)}\delta_{t}^{(a)},\hspace{0.3cm} e_{r}^{(a)}=\frac{1}{\sqrt{g(r)}}\delta_{r}^{(a)}, \hspace{0.3cm}e_{x}^{(a)}=r\delta_{x}^{(a)}.
\end{equation}
\begin{eqnarray}\label{metric_connections}\nonumber
&&\Gamma^{t}_{tr}=\frac{1}{\sqrt{g(r)}}\frac{d}{dr}\sqrt{g(r)},\hspace{0.3cm}\Gamma^{r}_{rr}=\sqrt{g(r)}\frac{d}{dr}\left(\frac{1}{\sqrt{g(r)}}\right),\\
&& \Gamma^{r}_{tt}=\frac{g(r)}{2}\frac{d}{dr}g(r),\hspace{0.3cm}\Gamma^{r}_{xx}=-rg(r),\hspace{0.3cm}\Gamma^{x}_{rx}=\frac{1}{r}.
\end{eqnarray}
From the above quantities it is possible to compute the components of spin connection (\ref{spin_connection}):
\begin{equation}\label{spin_components}
\omega_{t}^{(t)(r)}=\frac{1}{2}\frac{d}{dr}g(r),\hspace{0.3cm}\omega_{x}^{(r)(x)}=-\sqrt{g(r)}.
\end{equation}
Redefining the two component spinor $\Psi$ by
\begin{eqnarray}\nonumber
&&\Psi_1=i\left[r^2g(r)\right]^{1/4}e^{-i\omega t +i\kappa x}R_{+},\\
&&\Psi_2=\left[r^2g(r)\right]^{1/4}e^{-i\omega t +i\kappa x}R_{-},
\end{eqnarray}
the Dirac equation (\ref{dirac1}) becomes
\begin{equation}\label{dirac_2}
\left(\frac{d}{dr_{*}}\pm i\omega\right)R_{\pm}=WR_{\mp}.
\end{equation}
In the case of a massless spinor, e.g., a Weyl field, the superpotential $W$ is given by
\begin{equation}\label{superpotencial}
W=\kappa \frac{\sqrt{g(r)}}{r}.
\end{equation}
Now, letting $X_{\pm}=R_{+}\pm R_{-}$, we rewrite Eq.~(\ref{dirac_2}) as
\begin{equation}
\left(\frac{d^{2}}{dr_{*}^{2}}+\omega^2\right)X_{\pm}=V_{\pm}X_{\pm},
\end{equation}
where $V_{\pm}$ are the potentials for the massless two component spinor, 
\begin{equation}
V_{\pm}=W^{2}\pm \frac{dW}{dr_{*}}.
\end{equation}
With $W$ given by (\ref{superpotencial}), the potentials $V_{\pm}$ take the explicit form
\be
\label{w2}
V_\pm (r) = \frac{\kappa^2}{L^2}\bigg[ 1 - \left( \frac{r_+}{r} \right)^\sigma \bigg] \pm \frac{\kappa \sigma r_+^\sigma}{2L^3r^{\sigma -1 }} \bigg[ 1 - \left( \frac{r_+}{r} \right)^\sigma \bigg]^{1/2}.
\ee
As in the scalar field case, $\sigma$ plays the role of the quintessence charge, with $\sigma = 2+2w_q$. In Fig.~(5) various curves of $V_{\pm}(r)$ for different values of $w_q$ are presented. We are interested in the quasinormal frequencies obtained from this description when imposing the traditional boundary conditions in a typical AdS-like spacetime: incoming plane waves on the horizon and vanishing field at the AdS border, i.e., 
\be
\label{w3}
\lim_{r \rightarrow \infty} \Psi (r) & = & 0, \\
\label{w3b}
\lim_{r \rightarrow r_+} \Psi (r) & = & e^{-i \omega r_*}.
\ee
There are a few methods available to solve the Weyl field equation, and following the method of the previous section, we choose to expand the field in a manner similar to that of Ref.~\cite{Horowitz:1999jd}. Changing the radial coordinate to a more suitable one, $r^\sigma = y^{-1}$, in such a way that the horizon $r_+$ lies in $y_+$, we obtain a new field equation: 
\be
\nonumber
\Bigg\{ \sigma^2 y^{\sigma -1}\left(1 - \frac{y}{y_+} \right)^2\frac{\partial^2}{\partial y^2} + \Bigg[ y^{\frac{\sigma-2}{\sigma}}\left(1 - \frac{y}{y_+} \right)^2(\sigma^2-\sigma) \\
\label{w4}
- \frac{\sigma^2 y^{\sigma -1 }}{y_+}\left(1 - \frac{y}{y_+} \right)\Bigg] \frac{\partial}{\partial y} + \omega^2  - V_\pm (y)\Bigg\}\Psi(y) = 0,
\ee
with
\be
\label{w5}
V_\pm (y)= \frac{\kappa^2}{L^2}\left( 1 - \frac{y}{y_+}\right) \pm \frac{\kappa \sigma y^{\frac{\sigma - 1 }{\sigma}}}{2L^3 y_+}\left( 1 - \frac{y}{y_+}\right)^{1/2}.
\ee



\begin{figure}[htp!]\label{potenciais_weyl}
\centering
\includegraphics[scale=1.35]{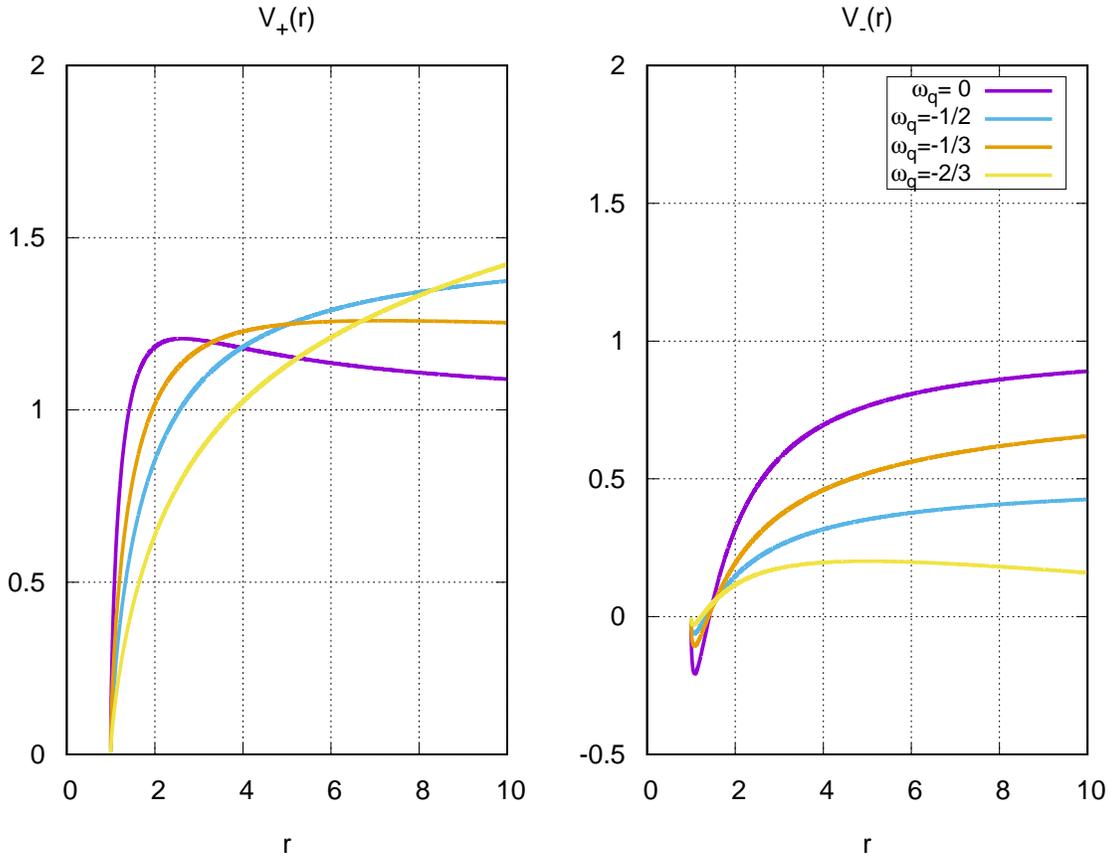}
\caption{Weyl field effective potentials for several values of $w_{q}$ with $M=L=\kappa=1$.}
\label{}
\end{figure}


Given the exponent $1/2$ in the second term of $V_\pm (r)$, we may choose as the ansatz an unusual expansion around the horizon, namely
\be
\label{w6}
\Psi (y) = \sum_{n=0}^{\infty}a_n(y-y_+)^{\alpha + \frac{n}{2}},
\ee
which seems to be the only way to couple with the half exponent in the effective potentials. Now, considering (\ref{w4}) in the form
\be
\label{w7}
\left[s(y)\frac{\partial^2}{\partial y^2 } + \tau  (y) \frac{\partial}{\partial y} + u(y) \right]\Psi(y)=0, 
\ee
we have the new functions $s, \tau$ and $u$ as expansions of the terms in (\ref{w4}),
\be
\nonumber
s(y) & = & \sum_{n=0}^{\infty}s_n(y-y_+)^n \hspace{0.3cm} \rightarrow \hspace{0.3cm} s_n= (0,0,s_2,s_3, s_4, \cdots ), \\
\nonumber
\tau(y) & = & \sum_{n=0}^{\infty}\tau_n(y-y_+)^n \hspace{0.3cm} \rightarrow \hspace{0.3cm} \tau_n= (0,\tau_1,\tau_2,\tau_3, \tau_4, \cdots ),\\
\label{w8}
u(y) & = & \sum_{n=0}^{\infty}u_n(y-y_+)^{\frac{n}{2}} \hspace{0.2cm} \rightarrow \hspace{0.2cm} u_n= (u_0,u_1,u_2,u_3, u_4, \cdots ).
\ee
Naturally, the series for $s, \tau$ and $u$ will depend on $\sigma$. In the cases with an integer quintessential exponent, we may have only one or two nonzero terms in $s_n$ and $\tau_n$, though an infinite in $u_n$. The noninteger $\sigma$ cases are quite more complicated, given that then $s_n$ and $\tau_n$ also have an infinite number of terms, but still converge with the chosen ansatz. 

From (\ref{w6}) and (\ref{w4}), we can obtain the two relevant relations. First, the indicial equation for $\alpha$, given by
\be
\label{w9}
\alpha (\alpha -1 ) s_2 + \alpha \tau_1 + u_0 = 0, 
\ee
which for every $\sigma$ reduces to  
\be
\label{w10}
\alpha = \pm \frac{i \omega (y_+)^{-\frac{1}{\sigma}}}{\sigma}.
\ee
The positive sign of $\alpha$ corresponds to incoming waves near the horizon and the negative stands for the outgoing waves, which are the proper quasinormal resonances in a similar fashion to \cite{Horowitz:1999jd}, which should be taken to solve the equation.

Second, the recurrence relation for $n>1$, with the chosen ansatz, reads
\be
\label{w11}
a_n = -\frac{1}{(n\alpha+n^2/4)\tau_1} \Bigg(\sum_{k=0}^{n-1} a_k u_{n-k} + \sum_{k=0}^{f(n)} a_l g_{n,k}\Bigg),
\ee
where $\tau_1=\sigma^2 y_+^{-1/\sigma}$,
\be
\label{w12}
f(n) =
\left\{ \begin{array}{c}
\frac{n-2}{2} \hspace{0.3cm} \rightarrow \hspace{0.3cm} n \hspace{0.1cm} \textrm{even}\\
\frac{n-3}{2} \hspace{0.3cm} \rightarrow \hspace{0.3cm} n \hspace{0.1cm} \textrm{odd}
\end{array} \right. , \hspace{0.5cm} 
l=
\left\{ \begin{array}{c}
2k \hspace{0.3cm} \rightarrow \hspace{0.3cm} n \hspace{0.1cm} \textrm{even}\\
2k+1 \hspace{0.3cm} \rightarrow \hspace{0.3cm} n \hspace{0.1cm} \textrm{odd}
\end{array} \right. ,
\ee
and
\be
g_{n,k} =
\left\{ \begin{array}{c}
(\alpha+k)(\alpha+k-1)s_{\frac{n+4}{2}-k} + (\alpha+k)\tau_{\frac{n+2}{2}-k}  \hspace{0.3cm} \rightarrow \hspace{0.3cm} n \hspace{0.1cm} \textrm{even}\\
(\alpha+k+1/2)(\alpha+k-1/2)s_{\frac{n+3}{2}-k} + (\alpha+k+1/2)\tau_{\frac{n+1}{2}-k}  \hspace{0.3cm} \rightarrow \hspace{0.3cm} n \hspace{0.1cm} \textrm{odd}
\end{array} \right. 
\label{w13}
\ee
In the above relations, we consider the first two terms of the expansion, $a_0=1$ and $a_1=-\frac{u_1}{(\alpha+1/4) \tau_1}$.

Now, if we truncate the series (\ref{w6}), in a particular number of terms, ($c)$, condition (\ref{w3}) requires that $\Psi(0)= \sum_n^c a_n (-y_+)^{n/2} = 0$, which is then the quasinormal equation we want to solve. The convergence of the method is tested in the usual way: after having obtained a particular mode for a given $c$, we repeat the procedure taking a larger number of terms (say, $c+10$) verifying whether the obtained frequencies are the same within a given precision. In general, for high $r_+$ and integer $\sigma$, this convergence is achieved with $c$ between 50 and 100 while for rational $\sigma$ around 150 terms\footnote{In a simple i7 processor, 8GB Ram, this would takes from 10 to 15 minutes.}. For small/intermediate $r_+$, however, the convergence is poor and other approaches, such as numerical integration are preferred. 

In Tables III and IV we list the outgoing quasinormal frequencies calculated via the Frobenius method as described above, for different values of $\sigma = 2(w_q +1)$, revealing that the modes for both potentials, $V_\pm$, are similar for large $r_+$, but show significant differences for small $r_+$. 

\begin{table}[htbp]
\centering
  \begin{tabular}{lSSSSSSSS}
    \hline
    \multirow{3}{*}{$r_{+}$} &
      \multicolumn{2}{c}{$w_{q}=0$} &
      \multicolumn{2}{c}{$w_{q}=-0.25$} &
      \multicolumn{2}{c}{$w_{q}=-0.333$} &
      \multicolumn{2}{c}{$w_{q}=-0.5$}\\
      \hline
      & {$\Re(\omega)$} & {$-\Im(\omega)$} & {$\Re(\omega)$} & {$-\Im(\omega)$} & {$\Re(\omega)$} & {$-\Im(\omega)$} & {$\Re(\omega)$} & {$-\Im(\omega)$} \\
      \hline
     2   & 1.0705 &2.1429  &0.7278 & 1.8173 & 0.6087 & 1.6849  & 0.3944 & 1.3693   \\
     10  & 0      & 6.0747    &0      & 4.5544 & 0      & 4.0512  & 0      & 3.0559   \\
     50  & 0      & 25.9563  &0      & 19.479 & 0      & 17.3207 & 0      & 13.0102  \\
     100 & 0     & 50.9441 &0      & 38.221 & 0      & 33.9803 & 0      & 25.5051  \\
     500 & 0     & 250.935 &0      & 188.214& 0      & 167.308 & 0      & 125.501  \\
    \hline
  \end{tabular}
  \caption{Weyl quasinormal frequencies for $L=\kappa=1$ for potential $V_+$.}
\label{Weyl_kappa1_vplus}
\end{table}

\begin{table}[htbp]
\centering
  \begin{tabular}{lSSSSSSSS}
    \hline
    \multirow{3}{*}{$r_{+}$} &
      \multicolumn{2}{c}{$w_{q}=0$} &
      \multicolumn{2}{c}{$w_{q}=-0.25$} &
      \multicolumn{2}{c}{$w_{q}=-0.333$} &
      \multicolumn{2}{c}{$w_{q}=-0.5$}\\
      \hline
      & {$\Re(\omega)$} & {$-\Im(\omega)$} & {$\Re(\omega)$} & {$-\Im(\omega)$} & {$\Re(\omega)$} & {$-\Im(\omega)$} & {$\Re(\omega)$} & {$-\Im(\omega)$} \\
      \hline
    0.2  & 0.9589 & 0.3574 & 0.9585 & 0.1582 & 0.7630 & 0.3778 & 0.7657 & 0.3431 \\
     2   & 0 & 0.4117  & 0 & 0.2905 & 0 & 0.2516  & 0 & 0.1749  \\
     10  & 0 & 4.1668  & 0 & 3.1071 & 0 & 2.7541  & 0 & 2.0456  \\
     50  & 0 & 24.088  & 0 & 18.053 & 0 & 16.0398 & 0 & 12.010  \\
     100 & 0 & 49.076  & 0 & 36.795 & 0 & 32.6999 & 0 & 24.505  \\
     500 & 0 & 249.067 & 0 & 186.789& 0 & 166.308 & 0 & 124.501 \\
    \hline
  \end{tabular}
  \caption{Weyl quasinormal frequencies for $L=\kappa=1$ for potential $V_-$.}
\label{Weyl_kappa1_vminus}
\end{table}

We check the accuracy of the values in Tab. \ref{Weyl_kappa1_vminus} and  \ref{Weyl_kappa1_vplus} by comparing the methods of Frobenius and numerical integration over null-coordinates, obtaining very similar results for both high $r_+$ (maximum error of $0.2\%$), and small $r_+$ (around $1\%$).  See Tab. \ref{Weyl_kappa1_comparison} for a specific example.

\begin{table}[htbp]
\centering
  \begin{tabular}{lSSSSSSSS}
    \hline
      \multicolumn{1}{c}{$r_+$} &
      \multicolumn{1}{c}{$\sigma=2$ (N)} &
      \multicolumn{1}{c}{$\sigma=2$ (F)} &
      \multicolumn{1}{c}{$\sigma=1$ (N)} &
      \multicolumn{1}{c}{$\sigma=1$ (F)}\\
\hline
2   & 0.4032$i$   &	0.4117$i$   &	0.1773$i$   &	0.1749$i$ \\
10  & 4.1325$i$   &	4.1668$i$   &	2.0511$i$   &	2.0456$i$ \\
50  & 24.0455$i$  &	24.0882$i$  &	12.0180$i$  &	12.0098$i$ \\
100 & 49.072$i$   &	49.076$i$   &	24.5163$i$  &	24.5050$i$ \\
500 & 249.095$i$  &	249.067$i$  &	124.568$i$  &	124.501$i$ \\
    \hline
  \end{tabular}
  \caption{Weyl quasinormal frequencies for $L=\kappa=1$ and potential $V_-$: comparison of numerical integration (N) with the use of prony method for the frequencies with the  Frobenius method (F).}
\label{Weyl_kappa1_comparison}
\end{table}
In general, the Frobenius method has a better convergence for large $r_+$ (as the expansion is done around $y_+^{-1}$)\cite{Horowitz:1999jd}, while the integration method is preferable for small $r_+$, as long as a typical grid scale for the profile acquisition is proportional to $1/r_+$; this imply a great computational cost when $r_+$ is high. 

From the above listed data, we can infer a scaling between $r_+$ and the quasinormal modes in the high-$r_+$ regime, 
\be
\label{w14}
\omega_{ \pm } = \left( \frac{r_+}{4} \pm \frac{\kappa}{2}  + O(r_+^{-1}) \right) \sigma i,
\ee
in which the different signs refer to the different potentials $V_\pm$. The scaling can be verified to happen also for other small $\kappa$ cases, such as $\kappa=3,4$, and is expected in the interpretation of $\Im(\omega )$ as a relaxation time in the AdS border.

In the Appendix, we obtain Eq.~(\ref{w14}) by means of an analytical solution in the case $\sigma = 1$ and address the same qualitative issue we quote in the previous section: the difference in the quasinormal spectrum for high and small black holes relative to the presence of $\Re (\omega)$. We demonstrate different border values for $r_+$ in which $\Re(\omega)>0$ for $V_\pm$.

\section{High-temperature scalar quasinormal frequencies}\label{hidrodinamicos}

In the last part of this work we will report on the presence of high-temperature modes, (also known as hydrodynamic approximation), by expanding the Klein-Gordon equation \red{(\ref{kg1})} in  the corresponding limit. Taking into consideration this equation in terms of the radial coordinate, we have 
\be
\label{h1}
\frac{\partial^2}{\partial r^2}\Psi + \left[ \frac{\partial_r h}{h} + \frac{1}{r} \right] \frac{\partial }{\partial r} \Psi + \frac{1}{h}\left[ \frac{\chi^2}{h} - \frac{\kappa^2}{r^2} \right] \Psi = 0,
\ee
where $h=r^2[1-(r_+ / r)^\sigma]/L^2$ is the $g_{00}$ metric term and $\chi$ is the time derivative of the field. Defining a change of variables $u=r_+ / r$, $f=1-u^\sigma$, and considering the Hawking temperature of the black hole, 
\be
\label{h2}
T_H = \frac{\sigma^2}{8\pi L^2} r_+,
\ee
we rewrite $\chi \rightarrow 8\pi T_H \omega / \sigma^2$ which, to leading order, substituting $\Psi \rightarrow f^\nu$ on (\ref{h1}), has only two different solutions:
\be
\label{h3}
\nu = \pm i \frac{\omega}{\sigma}.
\ee
Once again, we take as valid the solution of ingoing waves on the event horizon, thus, considering the negative signal in the above expression as the leading-order term in a near-horizon expansion for the high-temperature approach. In what follows this condition guides the procedure for the high-temperature modes expansion. We consider an ansatz of the form
\be
\label{h4}
\Psi = f^\nu (F_0 + i\omega F_1 - \omega^2 F_2 + \cdots ),
\ee
Substituting (\ref{h4}) into (\ref{h1}) yields
\be
\nonumber
\left[ - \frac{\omega^2}{\sigma^2}f^{\nu - 2}\dot{f}^2 + \frac{i \omega }{\sigma} f^{\nu - 2}\dot{f}^2 -\frac{i\omega}{\sigma}f^{\nu-1}\ddot{f} \right](F_0 + i\omega F_1) -\frac{2i\omega}{\sigma} f^{\nu -1 }\dot{f}(\dot{F_0} + i\omega \dot{F_1}) + f^\nu (\ddot{F_0} + i\omega \ddot{F_1})\\
+\left[ \frac{\dot{f}}{f} + \frac{\ddot{f}}{\dot{f}(1-\sigma )}\right] \left[ -\frac{i \omega }{\sigma }f^{\nu - 1} \dot{f} (F_0 + i\omega F_1 ) + f^\nu ( \dot{F_0} + i\omega \dot{F_1})\right] + \frac{\omega^2}{f^2}f^\nu ( F_0+i\omega F_1 ) =0, \hspace{0.9cm}
\label{h5}
\ee
which, organized in orders of $\omega$ gives
\begin{align}
\label{h6}
&\ddot{F_0} + \left[\frac{\dot{f}}{f} + \frac{\ddot{f}}{\dot{f}(1-\sigma )} \right] \dot{F_0}=0 &&\leftarrow (i\omega)^0, \\
\label{h7}
&\ddot{F_1} + \dot{F_1} \left[ \frac{\dot{f}}{f} + \frac{\ddot{f}}{\dot{f}(1-\sigma )}\right] - \frac{2\dot{f}}{\sigma f}\dot{F_0} + \left[ \frac{(\sigma -2)\dot{f}}{\sigma (1-\sigma ) f}\right] F_0=0 &&\leftarrow  (i\omega)^1.
\end{align}

The solution for Eq.~(\ref{h6}) is given in terms of hypergeometric functions of the first kind, written as

\be
\label{h8}
F_0 = A + B \frac{u^2}{2} {_2}F{_1} \left[1, \frac{2}{\sigma}, 1 +\frac{2}{\sigma}, u^\sigma \right],
\ee
where $A$ and $B$ are constants, which diverges $u \rightarrow 1$. Given the right boundary condition, we must take $B=0$, such that $F_0=A$ is the only allowed solution. Equation~(\ref{h7}) then leads to 
\be
\label{h9}
F_1 = C + D \frac{u^2}{2}\, {_2}F{_1} \left[1, \frac{2}{\sigma}, 1 +\frac{2}{\sigma}, u^\sigma \right]+ A \frac{\ln (1-u^\sigma)}{\sigma},
\ee
with $C$ and $D$ constants, which has two divergent functions for $u \rightarrow 1$. Taking a step back in the equation for $F_1$, we can put it into an integral form, 
\be
\label{h10}
F_1 = -A \int \frac{u^{\sigma -1}}{1-u^\sigma} du - D \int \frac{u}{1-u^\sigma}du + C.
\ee
In the near-horizon approximation, we can write the integrands expressions as,
\be
\label{h11}
\frac{u^{\sigma -1}}{1-u^\sigma} = -\frac{1}{\sigma}(u-1)^{-1} + \frac{1-\sigma}{2\sigma}(u-1)^0 + \frac{-5+6\sigma - \sigma^2}{12\sigma}(u-1)^1 + O(u-1)^2, \\
\label{h12}
\frac{u}{1-u^\sigma} = -\frac{1}{\sigma}(u-1)^{-1} + \frac{\sigma-3}{2\sigma}(u-1)^0 + \frac{-5+6\sigma - \sigma^2}{12\sigma}(u-1)^1 + O(u-1)^2.
\ee
We can then avoid divergences by imposing $A+D=0$, such that,
\be
\label{h13}
F_1 = C - A \frac{\sigma - 2}{\sigma}(u-1) + O(u-1)^3,
\ee
Taking this solution with $C=0$ and $F_0=A$ into the ansatz (\ref{h4}), we can write the wave function as 
\be
\label{h14}
\Psi = A(1-u^\sigma)^\nu \left[1+ i\omega \frac{\sigma -2}{\sigma}(1-u) \right],
\ee
which fulfills the quasinormal mode condition for ingoing waves in the horizon. Lastly, the Dirichlet condition of $\Psi (0) = 0$ in the border of the AdS spacetime yields for the high-temperature mode limit,
\be
\label{h15}
\omega=-i\frac{\sigma}{2-\sigma},
\ee
which represents a quasinormal mode in the cosmological range of interest for the variable $\sigma$: $0<\sigma < 2$, or, equivalently, $-1 < w_q < 0$. 

In the context of the AdS/CFT correspondence,  the dynamics of probe fields evolving in the bulk spacetime which contains a black hole are related to perturbations of a thermal state on the dual thermal field theory defined at the AdS boundary. Therefore the damping of the fundamental quasinormal frequency  of a scalar probe field in the bulk, with high  Hawking temperature, gives the characteristic thermalization time scale for the dual thermal state \cite{Horowitz:1999jd}. 

In our case, the time scale for the thermalization is 
\be\label{timescale}
\tau=\frac{1}{2\pi T_{H}}\left(\frac{2-\sigma}{\sigma}\right),
\ee
showing that, as for three-dimensional Lifshitz black holes \cite{Abdalla:2011fd}, the perturbation at the AdS  boundary is not long lived in the limit of high temperature. Notice that, as the quintessence parameter  $|w_q|$  increases, the time-scale $\tau$ for a given temperature $T_{H}$ increases as well demonstrating that the quintessence field in the bulk makes the return of the thermal state to equilibrium more difficult.

\section{Final remarks}\label{conclusoes}

In the present work, we studied a Schwarzschild-AdS-like black hole solution endowed with a quintessential field in $(2+1)$ dimensions, introduced in the line element via energy-momentum tensor. We have shown that the nature of the singularity depends on the quintessential charge $w_q$. We found two types of black holes: a spacelike singularity for $-1/2 < w_q \leq -1/3$ and a lightlike singularity for $w_q \leq -1/2$. This singularity is enclosed by a single event horizon and has an AdS spatial infinity, which allows us the study of interesting physical properties in the context of AdS/CFT conjecture. 

From this perspective, we investigated further the propagation of a scalar and a Weyl probe fields in the fixed geometry of the black hole. The scalar field profile in time domain had exhibit two different behaviors: a permanent damped oscillation or an exponential decay, depending on the spacetime parameters. 

The quasinormal spectra of probe scalar fields shows that, typically, higher $\sigma$ values (which correspond to lower $|w_q|$) lead to lower $\Re(\omega)$. For each quintessential charge $w_q$, $\Im(\Omega)$ and $r_+$ scale perfectly, supporting the interpretation of $\Im(\Omega)$ as relaxation time in the CFT context.

For the Weyl field perturbation we have a scale between the two quantities as well, of type $\omega_{I \pm} \propto \frac{\sigma r_+}{4} \pm \frac{\sigma \kappa}{2}$ for large $r_+$, which yields distinct spectra to second order in $r_+$, for $V_\pm$. Both scalar and Weyl fields display an oscillatory character only for low values of $r_+$, with the frequency being purely imaginary for high $r_+$. This fact will depend, however, on the $\kappa$ momentum of the field.

With respect to the distinct quasinormal spectra for the potentials $V_\pm$ of the Weyl case, we emphasize that the isospectrality property is related to the integral of a group of potential functions in the boarder of the spacetime \cite{Chandrasekhar:1985kt}. For instance, considering the superpotential $W$ in our Weyl case, we have
\be
\label{cc1}
\frac{d W}{d r_*} = \frac{d r}{d r_*} \frac{d W}{d r} = g_{tt}\frac{d W}{d r}= \frac{\kappa \sigma r_+^\sigma}{2L^3 r^{\sigma -1}} \bigg[ 1- \left( \frac{r_+}{r} \right)^\sigma \bigg]^{1/2}
\ee
and
\be
\label{cc2}
W= \frac{\kappa}{L}\bigg[ 1- \left( \frac{r_+}{r} \right)^\sigma \bigg]^{1/2},
\ee
leading to 
\be
\label{cc3}
W(r_+)=0; \hspace{1.0cm} W(\infty ) = \kappa.
\ee
Being $W|_{r_+}^{\infty}=0$ a necessary (but not sufficient) condition for isospectrality \cite{Cardoso:2003pj}, we obtained different spectra for the different potentials. This fact is reflected in the scale proposed as our result: the higher the $r_+$ and the smaller the $\kappa$, the more similar to each other the spectra produced by $V_+$ and $V_-$ are.

Finally, we also showed the presence of high-temperature quasinormal modes regarding a massless scalar field perturbation, and found that the quasinormal spectrum exhibits purely imaginary frequencies, as expected for a large AdS black hole. Considering, then, the AdS/CFT correspondence, we computed the thermalization time-scale $\tau$ of a thermal state on the dual field theory defined at the conformal AdS boundary, demonstrating the influence of the quintessence on the thermalization.

\begin{acknowledgments}
The authors would like to thank Jefferson Stafusa E. Portela and Alan B. Pavan for critical comments to the manuscript. This work was supported by CNPq (Conselho Nacional de Desenvolvimento Cient\'{\i}fico e Tecnol\'ogico), Brazil.
\end{acknowledgments}

\appendix* \label{apendice}
\section{Solving the field equations analytically: the case $\sigma =1$}

We are interested in the investigation of two specific aspects related to the quasinormal spectra showed in previous sections:  the qualitative change of purely imaginary oscillations to quasinormal modes with $\Re(\omega)>0$, and the absence of scale showed in the case $\sigma =2$. We will treat both fields with their specific potential and the variable transformation necessary to solve analytically the field equation.


\subsection*{Scalar field case}

We start with the scalar field equation in a tortoise coordinate system and the usual field transformation for $\sigma=1$,
\be
\label{e1}
\frac{d^2 \Psi}{dr_*^2} + \frac{r^2-rr_+}{L^2}\left[ \frac{-3+r_+/r}{4L^2}+\frac{\omega^2 L^2}{r^2 - r r_+}-\frac{\kappa^2}{r^2}\right]\Psi=0.
\ee
Changing the radial coordinate to $x=\frac{r-r_+}{r}$, an the field as $\Psi = (1-x)Z$ we have that 
\be
\label{e2}
x(1-x)Z'' + (1-3x)Z' + \left[ \frac{A}{1-x} + B - \frac{\delta^2}{x} + \alpha x \right] Z=0
\ee
in which the prime denotes derivative with respect to $x$, and the constants in the potential read $A=-3/4$, $B = A+\delta^2 - \alpha$, $\delta = i \omega L^2 r_+^{-1}$ and $\alpha = L^2 \kappa^2 r_+^{-2}$. Equation (\ref{e2}) can not be turned into a hypergometrical form, unless $\alpha =0$ \cite{nikiforov}. In such a case, we can find a field transformation, $Z= \chi u$, and the new equation is a hypergeometric differential equation. In our case, $\chi = (1-x)^{-3/2} x^{\delta}$ reduces (\ref{e2}) to 
\be
\label{e3}
x(1-x)u''+ (1+2\delta-2\delta x)u'+ \delta u=0,
\ee
which can be solved analytically in terms of a second-order hypergeometrical functions. A similar equation for $\sigma =2$ was found in ~\cite{Cardoso:2001hn}. There, the authors found two scales $Re(\omega) = \kappa$ and $\Im (\omega) \propto 2r_+$ which is not the case for other quintessential fields (see e. g. Eq.~(\ref{e7}) for $\alpha \neq 0$). The Dirichlet boundary condition reduces the quasinormal problem to the following relation,
\be
\label{e4}
\frac{-1+2\delta \pm \sqrt{1+4\delta^2}}{2} = n,
\ee
$n \in \mathbb{N}$. Then the quasinormal spectrum is given by
\be
\label{e5}
\omega = -\left( \frac{n^2+3n+2}{2n+3}\right) \frac{r_+}{L^2}i
\ee
whose quasinormal frequencies represents exactly what is displayed in Table~I (case $w_q=-1/2$), when $n=0$. 

Taking Eq.~(\ref{e2}) when $\alpha \neq 0$, we have
\be
\label{e6}
x(1-x)u''+ (1+2\delta-2\delta x)u'+ (\delta -\alpha (1-x))u=0,
\ee
or
\be
\label{e7}
u'' + \left(\frac{1}{1-x} + \frac{1+2\delta}{x}\right) u' + \left( \frac{\delta -\alpha}{x} + \frac{\delta }{1-x} \right) u=0.
\ee
which has a solution in terms of confluent Heun functions, expressed as
\be\nonumber
\label{e8}
u(x) = C_1 HeunC(0,2\delta , - 2, - \alpha , \alpha + 1, x) + C_2 x^{-2\delta}HeunC(0,-2\delta , - 2, - \alpha , \alpha + 1, x).\\
\ee
In terms of the first field variable, this solution turns to
\begin{eqnarray}\nonumber
\label{e9}
\Psi (x) = C_1 (1-x)^{-1/2}x^{\delta}HeunC(0,2\delta , - 2, - \alpha , \alpha + 1, x) \\
+ C_2 (1-x)^{-1/2}x^{-\delta}HeunC(0,-2\delta , - 2, - \alpha , \alpha + 1, x),
\end{eqnarray}
which diverges in the limit $x \rightarrow 1$ (or $r \rightarrow \infty$), unless $C_2 = 0$. 

The remnant confluent Heun function when expanded in a series of $x$ gives rise to $\Psi \rightarrow x^{\delta}$, when $r\rightarrow r_+$, which is the ingoing (to horizon) quasinormal wave. Applying the Dirichlet boundary conditions, i. e., taking $\Psi|_{x\rightarrow 1} = 0$, we have an equation of type
\be
\label{e10}
\Psi|_{x\rightarrow 1} \rightarrow \sum_n f_n (\delta , \alpha ) x^n \hspace{0.5cm} (=0 ),
\ee
where $f_n$ is the ratio of a polynomial of $\delta$ and $\alpha$ over a polynomial function of $\alpha$. The convergence of the series can be analyzed in a ‘pedestrial’ way, term by term, searching for a solution of a polynomial equation of type $\delta (\alpha )=0$. 

Being $\delta = iL^2 \omega r_+^{-1}$, we have a purely imaginary oscillation whenever $\delta \in \mathbb{R}$, and a $\Re(\omega)>0$ quasinormal mode if $\delta \in \mathbb{C}$. 
Taking an increasing number of terms in the series (\ref{e10}), the convergence of $\alpha$ is displayed in Table~\ref{ordens_escalar}.


\begin{table}[htbp!]
\centering
\begin{tabular}{|c|c|c|c|c|c|c|}
\hline
  $ O(1)$ & $ O(2)$ & $ O(3)$ & $ O(4)$ & $ O(8)$ & $O(14)$ & $O(18)$ \\
\hline
$ \forall \alpha, \Re(\omega)=0$ & $\alpha = 0.7927$ & $\alpha = 0.6801$ & $\alpha = 0.6646$ & $\alpha = 0.6564$  & $\alpha = 0.6550$ & $\alpha = 0.6549$ \\
\hline
\end{tabular}
   \caption{Solving the polynomial $\delta (\alpha ) =0$ for different orders in $x$.}
   \label{ordens_escalar}
\end{table}
The transition from $\Re(\omega) >0$ to purely imaginary modes is given by a critical value of $\alpha$, the highest order in the table,
\be
\label{e12}
\alpha_c \sim 0.6549,
\ee
such that 
\be
\label{e13}
r_+^{(c)} = \frac{\kappa L}{\sqrt\alpha_c}\sim 1.236 \kappa L.
\ee
Whenever $r_+<r_+^{(c)}$, we have $\Re(\omega) >0$, and if $r_+>r_+^{(c)}$, the quasinormal modes are purely imaginary. This is the reason why no transition is seen for $\kappa = 0$ in data from Sec.~III. The value of $r_+^{(c)} \sim 1.24$ was found also with numerical integration and with the Frobenius method (less than $1\%$ deviant), as the change in the spectrum from $\Re(\omega)=0$ to $\Re(\omega)>0$.

The same technique can be used to investigate the scale of $\delta$ and $\alpha$. Considering large black holes, or waves with small angular momentum, we can expand the polynomials $f_n$ for small $\alpha$ and solve the equation in $\delta$. The successive orders of (\ref{e10}) give rise to
\be
\label{e14}
\delta = -0.66706 - 0.37575\alpha - 0.11007\alpha^2 -O(\alpha^3) \hspace{0.3cm} \rightarrow \hspace{0.3cm} O(8), \\
\delta = -0.66681 - 0.37526\alpha - 0.11018\alpha^2 -O(\alpha^3) \hspace{0.3cm} \rightarrow \hspace{0.3cm} O(15),\\
\delta = -0.66676 - 0.37518\alpha - 0.11021\alpha^2 -O(\alpha^3) \hspace{0.3cm} \rightarrow \hspace{0.3cm} O(18),
\ee
with the following approximative result for the quasinormal frequencies,
\be
\label{e17}
\omega= -\left[ \frac{0.66676 r_+}{L^2} + 0.37518\frac{\kappa^2}{r_+} + 0.11021\frac{\kappa^4L^2}{r_+^3} + O\left(\frac{\kappa^6L^4}{r_+^5}\right) \right]i.
\ee
Expression (\ref{e17}) is compatible with Tables~\ref{scalar_qnm__k_0_hh} and \ref{scalar_qnm__k_1_hh} with an accuracy higher than $99.96\%$.

\subsection*{Weyl field case}

Taking the Weyl field equation with $\sigma =1$, a similar approach can be used. The wave equation expressed as
\be
\label{e18}
\left[ \frac{\partial^2}{\partial r_*^2} - \frac{\partial^2}{\partial t^2} - V(r) \right] \Psi(r,t)=0,
\ee
with 
\be
\label{e19}
V(r)= \frac{\kappa^2}{L^2}\left[1- \left(\frac{r_+}{r}\right) \right] \pm \frac{\kappa r_+}{2L^3 }\left[1- \left(\frac{r_+}{r}\right) \right]^{1/2}.
\ee
can be turned into a more suitable form by taking the variable $x^2 = 1- \frac{r_+}{r}$, given by
\be
\label{e21}
\left[ x^2 \frac{\partial^2}{\partial x^2} + x\frac{\partial}{\partial x} + (-a^2x^2 - bx + c^2) \right] \Psi=0,
\ee
in which
\be
\label{e22}
a= \pm \frac{2\kappa}{r_+}, \hspace{1.0cm} b=\pm \frac{2\kappa}{Lr_+}, \hspace{1.0cm}
c=\frac{2\omega L}{r_+}.
\ee
Equation (\ref{e21}) can be analytically solved when $L=1$, ($a= b$), in terms of modified Bessel functions \cite{arfken}, 
\be
\nonumber
\Psi = C_1 \sqrt{x}\left[ Bessel I \left( -\frac{1}{2}+ic , ax \right) + Bessel I \left( \frac{1}{2}+ic , ax \right)\right] + \\
\label{e23}
C_2 \sqrt{x} \left[ Bessel K \left( -\frac{1}{2}+ic , ax \right) +Bessel K \left( \frac{1}{2}+ic , ax \right) \right]
\ee
Solution (\ref{e23}) can be used for both $V_+$ and $V_-$, adjusting the signal of $b$ ($a=b$). The ingoing wave is represented by the solution with $C_1$ (while $C_2$ represents an outgoing, being discarded). The Dirichlet condition in spatial infinity reads $\Psi |_{x=1}=0$, which can be expressed in exactly the same way of Eq.~(\ref{e10}). In a similar fashion, we can solve an equation $c(b) =0$ and test the limits of $c$ purely imaginary or with $\Re(c)>0$. Performing it for both potentials, up to $O(40)$, a critical $b$ arises for each case,
\be
\label{e24}
b^{(c)} = 2/\alpha_c^\pm, 
\ee
with $\alpha_c = (r_+ /\kappa)^{(c)}$,
\be
\label{e25}
\alpha_c^{+} \sim 2.720, \\
\label{e26}
\alpha_c^{-} \sim 1.111. 
\ee
The scale for which we have purely imaginary modes varies with the potential (what can be seen in Tables~\ref{Weyl_kappa1_vplus} and \ref{Weyl_kappa1_vminus}): for $V_+$, we have $\Re(\omega)=0$ if $r_+>2.72 k$, while for $V_-$ smaller black holes ($r_+>1.111 k$) have also $\Re(\omega)=0$.

Expression $(\ref{w14})$ can be obtained as a particular case of the procedure used above. The expansion of $c(b) =0$ for small $b$ produces a solution of type 
\be
\label{e27}
c=-\left[ \frac{1}{2} + \frac{b}{2} + \frac{b^2}{4} + \frac{b^3}{8} + O(4) \right]i,
\ee
which scales the quasinormal mode as
\be
\label{e28}
\omega =-\left[ \frac{1}{2} \pm \frac{\kappa}{r_+} + \frac{\kappa^2}{r_+^2} \pm \frac{\kappa^3}{r_+^3} + O(4) \right]\frac{r_+}{2} i,
\ee
the first two terms being exactly those of (\ref{w14}).

\section*{References}

\bibliography{referencias}

\end{document}